\documentclass[]{aa}   

\usepackage{graphicx}
\usepackage{txfonts}
\usepackage{xcolor}
\usepackage{caption}
\usepackage{subcaption}
\usepackage{ulem}
\newcommand{\Rstart}{R$_{start}$}

\newcommand{\fdust}{f$_{dust}$}
\newcommand{\fplanet}{f$_{planetesimal}$}
\newcommand{\fcarbon}{f$_{carbon}$}

\begin{document}

   \title{Constraining the formation of WASP-39b using JWST transit spectroscopy}

   \author{N. Khorshid
          \inst{1}
          \and
          M. Min
          \inst{2}
          \and
          J. Polman
          \inst{3}
          \and
          L.B.F.M. Waters
          \inst{1,4}
          }

    \institute{Department of Space, Earth and Environment, Chalmers University of Technology, Gothenburg SE-412 96, Sweden\\
             \email{niloofar.khorshid@chalmers.se}\\
         \and
            SRON Leiden, Niels Bohrweg 4, 2333 CA Leiden, the Netherlands\\
        \and
            Space Research and Planetary Sciences, Physics Institute, University of Bern, Gesellschaftsstrasse 6, 3012 Bern, Switzerland\\
        \and
            Department of Astrophysics/IMAPP, Radboud University Nijmegen, PO Box 9010, NL-6500 GL Nijmegen, the Netherlands\\}

   \date{Received Month 00, 2021; accepted Month 00, 2021}

  \abstract
   {Understanding the formation history of planets is one of the goals of studying exoplanet atmospheres. The atmospheric composition of planets can provide insights into the formation pathways of planets. Even though the mapping of the atmospheric composition onto a formation pathway is not unambiguous, with the increasing sensitivity of modern instruments, we can derive promising constraints.}
   {In this work, we aim to understand the formation pathway of WASP-39b. We discuss whether the detection of SO$_2$ in its atmosphere would impact our understanding of the formation of the planet and whether it enables us to determine the formation pathway of the planet with greater accuracy.}
   {We used the JWST transit observation of the planet together with the available HST and Spitzer observations. We used a formation model coupled with a radiative transfer retrieval model to derive the planet's atmospheric characteristics and formation history. Furthermore, we used a photochemical model to derive the impact of photochemistry on the atmosphere of the planet.}
   {In this work, we show that the planet is most likely to have initiated beyond the CO$_2$ ice line of its natal disk. Furthermore, the planet is likely to have have accreted some planetesimals during its formation. We show that the sulfur abundance in the atmosphere of the planet is probably lower than $2.27 \times 10^{-4}$. This abundance indicates that the planet is likely to exhibit a lower metallicity than suggested by the retrievals. Furthermore, such an abundance for sulfur is more likely if WASP-39b had been formed beyond the CO ice line of its natal disk.}
   {}

   \keywords{planet formation --
                planet atmosphere
               }

   \maketitle

\section{Introduction}
\label{sec: introduction}

Studies on planet formation suggest there is a connection between planetary composition and their formation pathways. \citet{Oberg_2011} have shown that the location where the planet is formed with regard to the disk ice lines determines the carbon over oxygen, C/O, ratio of planets. Additionally, formation models suggest a connection between the metallicity of the planets and the C/O ratios related to their atmospheric enrichment \citep{Schneider_2021}. The planetary C/O ratio can provide information about the migration path during the formation of the planets \citep{Madhusudhan_2014, Shibata_2022}. Furthermore, the abundance of heavy elements in the atmosphere of gas giants may contribute to constraining the formation location of the planet \citep{Chachan_2023}. 

Another tracer of the formation pathway of planets is the nitrogen over oxygen ratio \citep{Cridland_2020_n2, Notsu_2022}. The nitrogen abundance particularly is believed to be relevant to the formation of Jupiter \citep{Ohno_2021, Bosman_2019}. Refractory elements such as magnesium, or sodium are other tracers to determine the amount of solid accretion during the formation of the planet as these elements condense to their solid forms very close to the host star \citep{Turrini_2020,Oberg_2021}. 

The atmospheric composition is not only able to provide information on the formation history of the planets, but it can also serve as an indication of the evolution of the composition of the natal disk \citep{pacetti_2022}. All these studies point out the fact that the planetary composition would allow us to peek into their formation history. 
\citet{Khorshid_2022a} have shown a correlation between the metallicity and C/O ratio of the planets and their initial formation locations. In the same study, they discuss the fact that this correlation offers a tighter constraint when the planet has a super-solar metallicity. 

The \textit{James Webb} Space Telescope (JWST) is capable of measuring the chemical composition of exoplanet atmospheres with unprecedented accuracy, allowing for more precise comparisons with predictions from planet formation models. Here, we focus on the Saturn-mass planet WASP-39b. Previous studies of the atmospheric composition of WASP-39b showed that the planet has a super-solar metallicity \citep{Wakeford_2018, Thorngren_2019}. The recent results obtained from JWST observations suggest a metallicity of around 10$\times$solar \citep{Rustamkulov_2023}. These observations offer the highest level of accuracy we have been able to achieve to date, which makes WASP-39b a prime target with respect to the study of its formation.

Observations carried out by JWST also suggest the presence of SO$_2$ in the atmosphere of WASP-39b \citep{Alderson_2023,Rustamkulov_2023,Tsai_2022}. Sulfur is expected to be a good tracer of the formation history of planets because of the refractory nature of sulfur in planet-forming disks, making it a useful tracer of metallicity \citep{legal2021,kama2019}. Sulfur bearing molecules are expected to be observable, especially in planets with high metallicities and low C/O ratios \citep{Polman_2022}. Given the high temperatures in its atmosphere, the atmospheric sulfur content of WASP-39b is expected to be in the gas phase \citep{carone_2023}.

In this work, we aim to constrain the formation history of WASP-39b using available observations of its atmosphere. In addition to the C/O ratio and metallicity of the planet, we use the strength of the SO$_2$ feature. This feature has been previously reported in \citet{Tsai_2022} and is expected to be due to photochemistry-induced reactions in the atmosphere of WASP-39b. 

In Section \ref{sec: method}, we describe the steps we undertook towards this goal. Section \ref{sec: results} presents over the main results of this study. In Section \ref{sec: discuss}, we discuss the results of this work, assumptions, and limitations. Finally in Section \ref{sec: conclusion}, we conclude with our main results.

 \section{Method}
\label{sec: method}

In this study, we used the James Web Space Telescope (JWST) observation from the Early Release Science program using the prism mode of NIRSpec instrument on WASP-39b \citep{Rustamkulov_2023}. Since the different data reduction methods used on the NIRSpec data provided very similar results we, rather arbitrarily, picked the data that was cleaned using FIREFLy \citep{NIRCam}. We combined this observation with the observations done by the Hubble Space Telescope (HST) and Spitzer of WASP-39b from 0.3 to 6 micrometers reported in \citet{Wakeford_2018}. These data were put together from different observations that had been used in \citet{Fischer_2016, Sing_2016, Wakeford_2018}. The HST and JWST spectra were obtained using different data reduction pipelines, which might lead to slight systematic offsets in the transit spectrum. To account for this, we fitted for a scaling factor of the JWST spectrum. This factor is added as a retrieval parameter.

In addition to the accurate measurement of the atmospheric metallicity of WASP-39b, the detection of the SO$_2$ feature in the spectra of the planet, observed with JWST, motivated us to study the formation path of WASP-39b. The use of sulfur as a tracer of the formation of planets was previously studied in \citet{Polman_2022}. Furthermore, the planet formation model we used enables us to measure the abundance of sulfur that is accreted during the formation of the planets. 

We acknowledge that the best method to study sulfur abundance in the atmosphere of WASP-39b would be using an atmospheric retrieval that includes photochemistry. However, the large computational cost of photochemical models makes their incorporation into retrieval models computationally impractical. Instead, the method that is used in this study consists of an atmospheric retrieval using equilibrium chemistry followed by a forward model that uses the results of the retrieval and makes use of photochemistry. This method is further explained in Section \ref{sec1: method-photochem}.

To retrieve the atmospheric parameters of WASP-39b we used the atmospheric modeling and retrieval code ARCiS (ARtful modeling Code for exoplanet Science) \citep{Min_2020}. To retrieve the formation path that would result in an atmospheric composition compatible with the atmospheric observation of WASP-39b, we used the formalism presented in \cite{Khorshid_2022a} and implemented in the SimAb code. We then used VULCAN to simulate the photochemistry in the atmosphere of WASP-39b \citep{Tsai_2021} and ARCiS to model the atmosphere of the planet. Each element of the methodology is detailed below.

\subsection{Atmospheric retrieval}
\label{sec1: method-ARCiS}

We used the atmospheric modeling and retrieval code ARCiS to retrieve the planet parameters of WASP-39b from the observed transit spectrum. We used the Multinest algorithm \citep{Feroz_2009} to sample the parameter space and obtain the posterior distribution. For this study, we used the implemented cloud formation model that is described in \citet{Ormel_2019} together with the equilibrium chemistry module GGchem \citep{Woitke_2018} and planet formation module, SimAb \citep[][explained below]{Khorshid_2022a}. We used the same retrieval setup as was used in \citet{Min_2020} but for the planet formation retrievals we replaced the parameters describing the atomic composition with a module computing the atomic abundances from planet formation (as described in the next section). The cloud formation model was adjusted to include Na/K silicates in addition to the Mg/Fe silicates that were initially used in \citet{Min_2020}.

\subsection{Formation model}
\label{sec1: method-SimAb}
To retrieve the formation parameters of WASP-39b, we used the planet formation formalism as explained in \citet{Khorshid_2022a} (from hereon referred to as Paper I) and implemented in the SimAb code which is a fast and flexible planet formation simulation. By parameterizing the physics involved in planet formation, SimAb enables studies of the impact of different formation processes on the total atomic abundance of the atmosphere of planets. SimAb models the part of the formation of gas giants where the planet is migrating while accreting gas and solid. This model assumes the atmosphere of the planet is accreted in a phase where the planet is more massive than the pebble isolation mass; therefore, the only solid-size material that can be accreted at this phase are either planetesimals and or dust grains. In this simulation, the planetesimals are assumed to be captured in the feeding zone of the planet while the dust grains are coupled to the gas and accreted with the gas. SimAb assumes the chemistry of the disk does not evolve as the planet is forming. Furthermore, it assumes that the total atomic abundance of the natal disk of WASP-39b is solar which is similar to its host star \citep{Polanski_2022}

The formation processes were simulated using four main parameters. The dust fraction, \fdust, describes the fraction of solid particles that have a Stokes number much less than 1 and that are coupled to the gas. The planetesimal ratio, \fplanet, is interpreted as the product of the fraction of solid material in planetesimals and the efficiency with which these are accreted. The last parameter is the carbon fraction at the soot line, \fcarbon. This parameter determines the amount of carbon that stays in the solid phase at temperatures below 800K and affects the carbon and oxygen fraction through out the disk. Finally, the migration initial distance, \Rstart, determines the distance from where the planet initiated its Type II migration. Since we showed in paper I that the initial core mass is not correlated with the atomic abundance of the atmosphere, in this study, the core mass was fixed to ten Earth masses. For more details on SimAb, we refer to Paper I.

\subsection{Retrieval setup}
\label{sec1: method-integrated}
For this study, we integrated SimAb in ARCiS to study the formation pathway of WASP-39b. Thus, SimAb replaces the way that the elemental abundances are calculated in ARCiS. In this mode, an extra step is added to ARCIS, where the atmospheric composition is calculated through SimAb for every given formation parameter. These calculated atmospheric parameters are used by ARCiS. Therefore, instead of the atomic ratios (e.g., C/O, Si/O, and others) that were the retrieved parameters previously, these are now replaced by the formation parameters described above. To obtain the derived posterior distribution of the atomic ratios in this approach, the posterior distribution of the retrieved formation parameters were plugged into SimAb formation simulations, which provide a range of chemical compositions resulting from these retrieved formation parameters.

In addition to these parameters describing the atomic composition, we retrieve the nucleation rate ($\dot{\Sigma}$) and the diffusion coefficient (K$_{zz}$) that describe the cloud formation in the atmosphere. The temperature structure is parameterized using the parameterization from \citet{Guillot_2010}. The parameters for the integrated retrieval used in this study and their priors are provided in Table \ref{tab: param-integrated}.

\begin{table*}[ht!]
                \begin{center}
                \caption{Retrieval parameters in the integrated retrieval}
                        \begin{tabular}{ |c|c|c|c| } 
                                \hline
                                Retrieval parameters & Symbol & Range & Prior\\
                                \hline
                                Planet radius& R &$5\sigma$ around lit. value&Flat linear\\
                                Ratio of visible to IR opacities& $\gamma$ &$0.01$ ... $100$ &Flat log\\
                                Cosine of the angle of irradiation&$\beta$& $0.0 .. 0.25$&Flat linear\\
                                IR opacity&$\kappa_{IR}$&$10^{-4}$ ... $10^{4}$&Flat log\\
                                Internal temperature of planet&T$_{int}$&$10$ ... $3000$ &Flat log\\
                                log(g) of planet&$log(g)$&$5\sigma$ around lit. value& Gaussian prior\\
                                Diffusion coefficient &$k_{zz}$&$10^{5}$ ... $10^{12}$&Flat log\\
                                Nucleation rate& $\dot{\Sigma}$&$10^{-17}$ ... $10^{-7}$&Flat log\\
                                Initial migration distance&\Rstart& Planet location ... $200$ (AU)&Flat linear\\
                                planetesimal ratio&\fplanet&$0$ ... $1$&Flat linear\\
                                dust grain fraction&\fdust&$0$ ... $1$&Flat linear\\
                                carbon fraction&\fcarbon&$0$ ... $1$&Flat linear\\
                S/H ratio\footnotemark & S/H & $0$ ... $2.27\times 10^{-4}$ & Flat Linear\\
                                \hline
                        \end{tabular}
                        \label{tab: param-integrated}
                \end{center}
                
        \end{table*}
\footnotetext{This prior is only included to retrieve the formation parameters that correspond to an atmospheric S/H ratio of $2.27\times 10^{-4}$. These results throughout this paper are shown via the blue color.}
The above setup assumes the observed atmosphere is unaltered after formation and mechanisms such as atmosphere evaporation have not influenced the total atomic abundance of the atmosphere. We return to this topic in the discussion in Section 4.

Besides the integrated formation retrieval we also performed retrievals using the exact same setup as in \cite{Min_2020}, namely, using the elemental abundance ratios as retrieval parameters. We refer to this setup as "free retrievals." We refer to the setup where the formation model is integrated in the retrieval "integrated retrievals." The parameters for free retrieval used in this study and their priors are provided in Table \ref{tab: param-free}.

\begin{table*}[ht!]
                \begin{center}
                        \caption{Retrieval parameters in the atmospheric free retrieval}
                        \begin{tabular}{ |c|c|c|c| } 
                                \hline
                                Retrieval parameters & Symbol&Range & Prior\\
                                \hline
                                Planet radius& R &$5\sigma$ around lit. value &Flat linear\\
                                Ratio of visible to IR opacities& $\gamma$&$0.01$ ... $100$ &Flat log\\
                                Cosine of the angle of irradiation&$\beta$&$0.0$ ... $0.25$&Flat linear\\
                                IR opacity&$\kappa_{IR}$&$10^{-4}$ ... $10^{4}$&Flat log\\
                                Internal temperature of planet&T$_{int}$&$10$ ... $3000$ &Flat log\\
                                log(g) of planet&$log(g)$& $5\sigma$ around lit. value& Gaussian prior \\
                                Diffusion coefficient &$k_{zz}$&$10^{5}$ ... $10^{12}$&Flat log\\
                                Nucleation rate& $\dot{\Sigma}$&$10^{-17}$ ... $10^{-7}$&Flat log\\
                    C/O ratio&C/O&$0.1$ ... $1.3$&Flat linear\\
                                Si/O ratio&Si/O&$0$ ... $0.3$&Flat linear\\
                                N/O ratio&N/O&$0$ ... $0.3$&Flat linear\\
                                Metallicity&[Z]&$-1$ ... $3$&Flat linear\\
                                \hline
                        \end{tabular}
                        \label{tab: param-free}
                \end{center}
                
        \end{table*}

 To derive the retrieval parameters, ARCiS uses MultiNest \citep{Feroz_2009}. The priors given in Table \ref{tab: param-integrated} and \ref{tab: param-free} were used to calculate the likelihood of the models and find the parameter sets that are best at reproducing the observed spectra. Even though S/H is not one of the retrieved parameters, we performed a third retrieval with a strict prior on the S/H value to prefer those models that have a specific sulfur content. This prior is a step function that is zero everywhere that the S/H is higher than the desired value and is one everywhere that is lower than the desired value. This further constraint allows for studying models that have enough sulfur abundance to reproduce a SO$_2$ feature similar to what is observed for WASP-39b.

\subsection{Photo-chemistry}
\label{sec1: method-photochem}

The atmospheric retrieval code used in this study does not include photochemistry. Nevertheless, we realize that this is an important effect in adjusting the molecular content of the atmosphere.
The study by \citet{Tsai_2022} have shown that the feature at around 4 microns could be due to the SO$_2$ produced through the photo-dissociation of water, which then reacts with sulfur to form SO$_2$. They have shown that the strength of the SO$_2$ feature is related to the atmospheric metallicity. Therefore, we used the VULCAN code to simulate the chemical reactions in the atmosphere of WASP-39b, while including the effects of photo-chemistry as a post-processing step after the retrieval is completed.

We used the pressure-temperature profile and elemental composition of some of the models from the retrieved posterior distribution using the integrated ARCiS-SimAb model. We chose models that have various S/O ratios and S/H ratios. Within VULCAN, we used the radius of WASP-39, 0.895 R$_\odot$. For the planetary parameters, we used an orbital radius of 0.486 AU, a planetary radius of 1.27 R$_J$, and a surface gravity of 426 cm/s$^2$, based on the data of \citet{Faedi_2011}. We then used ARCiS to model the spectra for these planets given the mixing ratios of molecules that VULCAN had calculated.
The results of the forward model were then used to add an extra prior on the S/H ratio, as explained in Section \ref{sec1: method-integrated}. This allows us to inspect whether including the impacts of photochemistry on the atmosphere of WASP-39b changes our understanding of the formation history of the planet.

\section{Results}
\label{sec: results}
\subsection{Comparing free atmospheric retrieval and integrated atmospheric retrieval}

\begin{figure}[t!]
\centerline{\resizebox{\hsize}{!}{\includegraphics{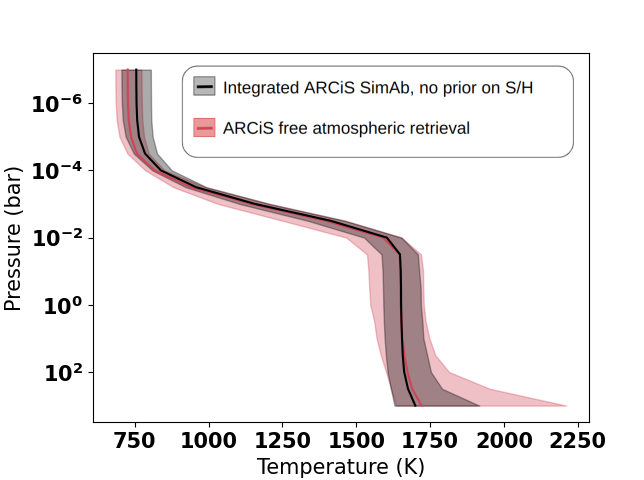}}}
        \caption{Retrieved temperature-pressure profile of WASP-39b when using the ARCiS free retrieval in red and when using Integrated ARCiS/SimAb in black.}
        \label{P3-fig: ptstructure}
\end{figure}

\begin{figure}[t!]
\centerline{\resizebox{0.9\hsize}{!}{\includegraphics{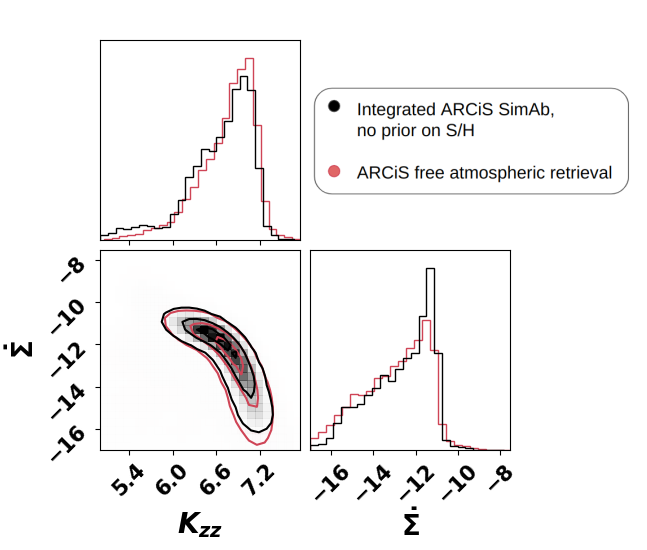}}}
        \caption{Comparison between the cloud forming parameters when using ARCiS free retrieval (red) to those parameters when using the integrated ARCiS/Simab}
        \label{P3-fig: atmosphere}
\end{figure}

Figures \ref{P3-fig: ptstructure} and \ref{P3-fig: atmosphere} show the atmospheric retrieval comparison between the free atmospheric retrieval and the integrated formation-atmospheric retrieval. These figures demonstrate that the temperature-pressure profile and the cloud formation parameters from both methods are similar.

\begin{figure}
    \centering
    \includegraphics[height=8cm]{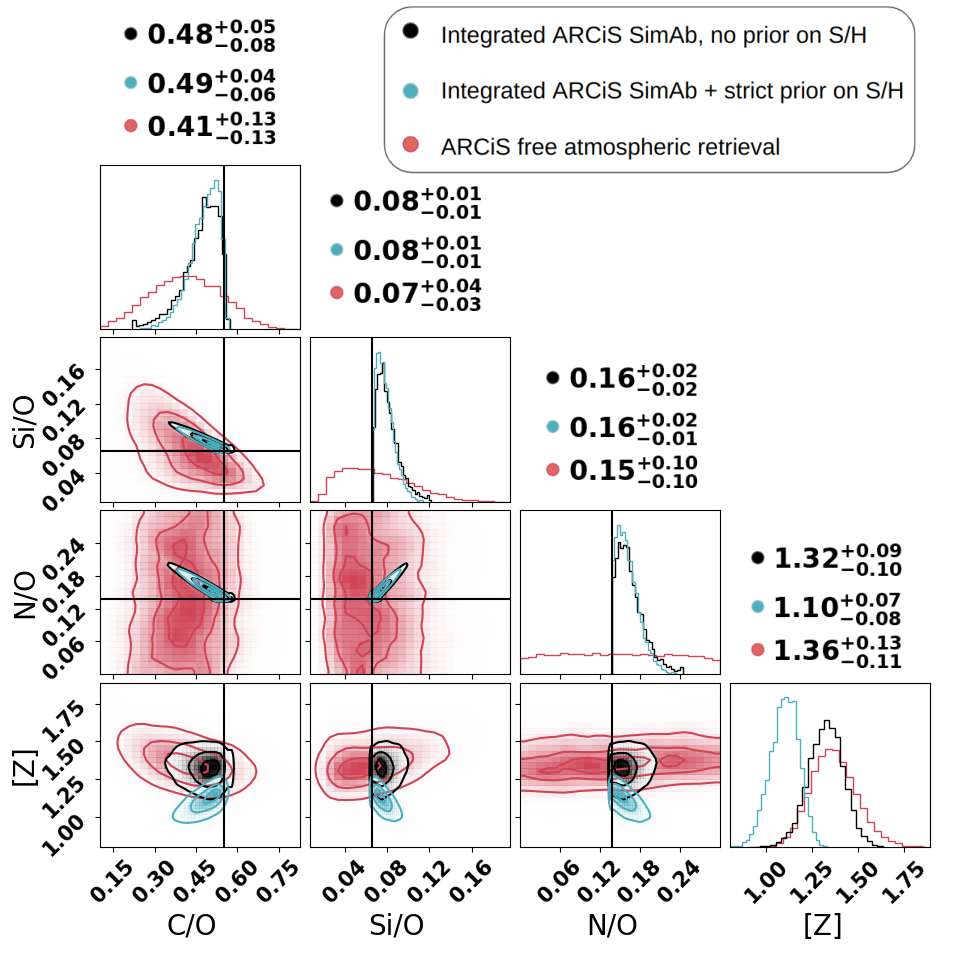}
    \caption{Retrieved ratios of some key elemental abundances in the atmosphere of WASP-39b using ARCiS free retrieval in red compared to the results when using the integrated ARCiS/Simab in black. The blue distribution shows these ratios when derived using a strict prior for the S/H and using integrated ARCiS/SimAb retrieval. The solid lines indicate the solar values.}
    \label{P3_fig: comp}
\end{figure}

To compare the free and integrated retrievals, we compute the atomic abundance ratios from the posterior distribution of the formation parameters obtained from the integrated retrievals. When comparing the elemental abundance in the atmosphere between the two methods, the free atmospheric retrieval results in a less constrained results overall. Figure \ref{P3_fig: comp} shows a comparison between the ratio of the abundance of some key elements in the atmosphere of WASP-39b that is retrieved through free atmospheric retrieval and is derived from the retrieved formation parameters through integrated retrieval. This figure shows that aside from the metallicity of the atmosphere, all the other parameters, C/O, Si/O, and N/O are less restricted when obtained from the free atmospheric retrieval compared to the integrated retrieval method. We conclude that the restriction on the ratio of the elemental abundance in the integrated retrieval method is due to the prior imposed by SimAb on the elemental abundance of planetary atmosphere.

\subsection{Formation retrievals}

\begin{figure*}
    \centering
    \includegraphics[height=13cm]{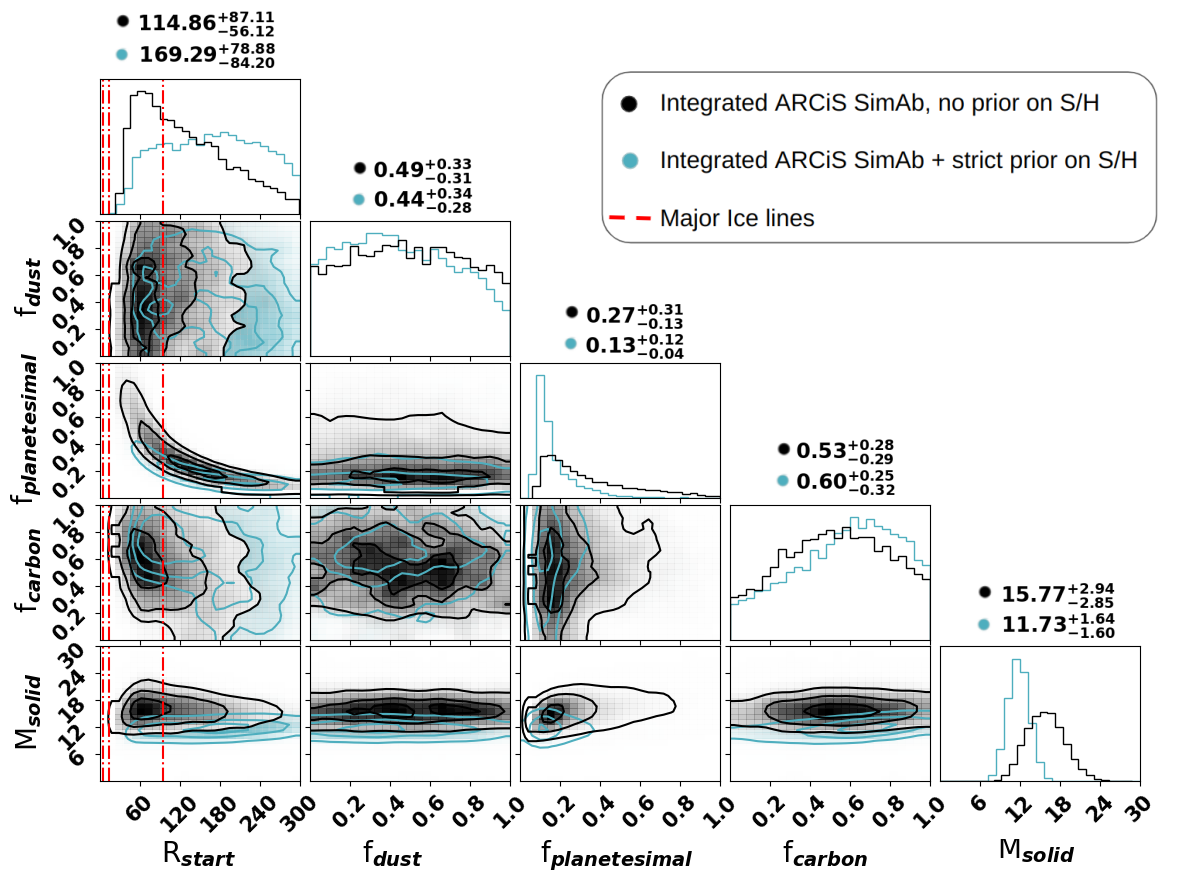}
    \caption{Retrieved formation parameters of WASP-39b, using integrated ARCiS/SimAb retrieval. The black distributions show the results when there is no prior on the sulfur abundance. The blue distributions show the results when there is a prior on the sulfur abundance.}
    \label{P3_fig: form}
\end{figure*}

When using the integrated retrieval method, the retrieval gives the possible formation parameters for WASP-39b directly. The black contours in Figure (\ref{P3_fig: form}) show the retrieved formation parameters of WASP-39b.
This figure shows that there is no constraint on the dust fraction in the disk. There is a slight preference for the carbon fraction at the soot line to be above 40 percent. However, given the high uncertainty, we cannot draw any conclusion from this value. As was discussed in paper I, the most constrained parameters are the planetesimal ratio and the migration distance with lower limits of $0.14$ for the planetesimal ratio and $58.74$ AU for the orbital distance from where the planet initiates its migration. This orbital distance is beyond the CO$_2$ ice line which is at 12.83 AU. Therefore, this result shows that WASP-39b should have initiated its Type II migration beyond the CO$_2$ ice line. It is important to point out that the luminosity and the mass of the host star influence the ice line locations. Table (\ref{tab:iceline}) presents the temperature and the location of the main ice lines in this model for a host star with mass and luminosity of WASP-39. Furthermore, Figure (\ref{P3_fig: form}) shows that there is a degeneracy between the migration distance and the planetesimal ratio. This degeneracy is caused by the fact that in our model more planetesimals can be accreted by either increasing the planetesimal fraction or migrating over a larger distance (allowing to sweep a larger part of the disk clean of planetesimals). Either of these two options can create a high metallicity and the retrieval cannot discriminate between them.

\begin{table}[]
    \centering
    \caption{Locations of the major ice lines.}
    \begin{tabular}{|c|c|c|}
    \hline
         Ice & Temperature (K)& Distance (AU)\\
    \hline 
    H$_2$O& 120& 4.32\\
    CO$_2$&47&12.83\\
    CO&20&94.19\\
    \hline
    \end{tabular}
    \label{tab:iceline}
\end{table}

\subsection{Sulfur abundance}
\label{sec1: S-abun}
\begin{figure*}
\centerline{\resizebox{\hsize}{!}{\includegraphics{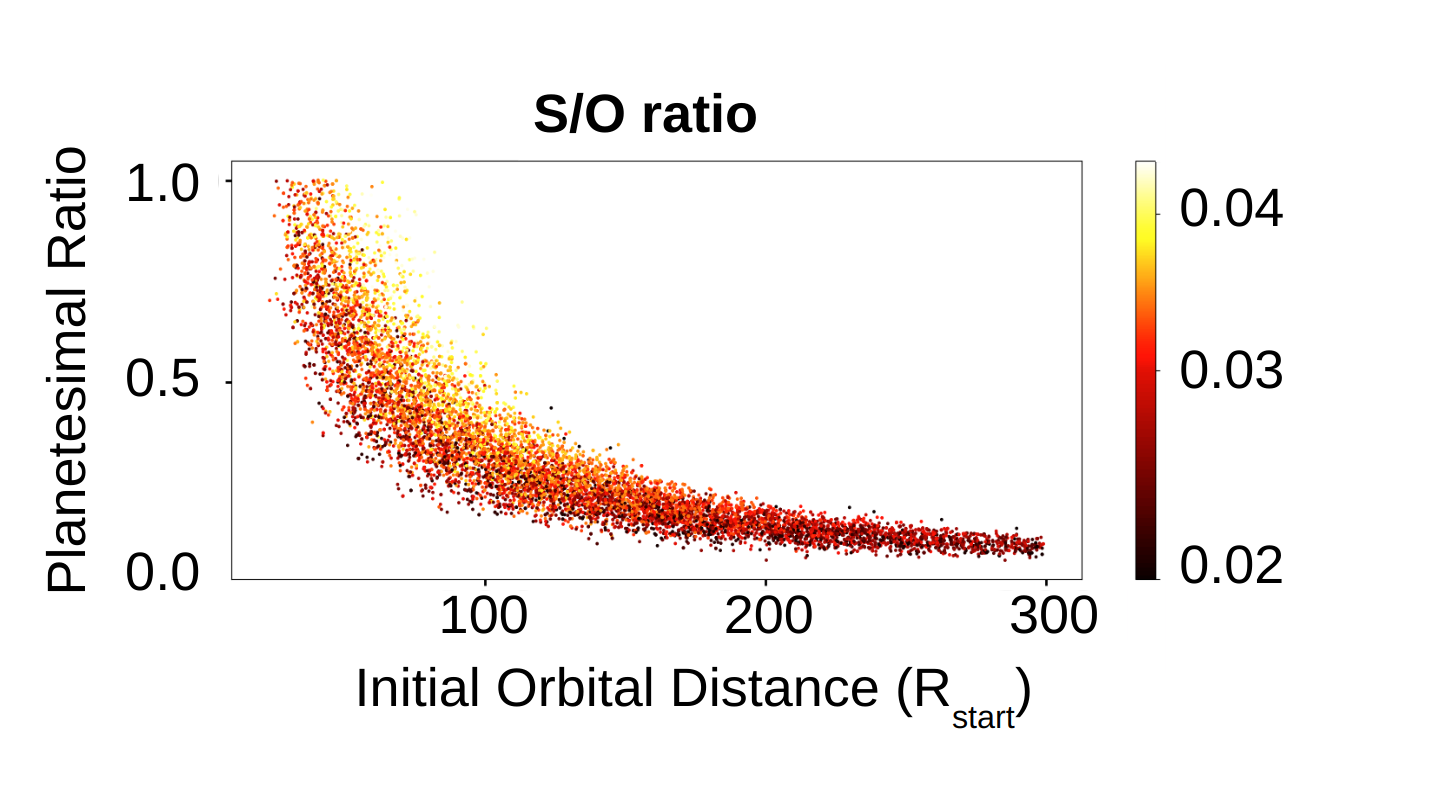}\includegraphics{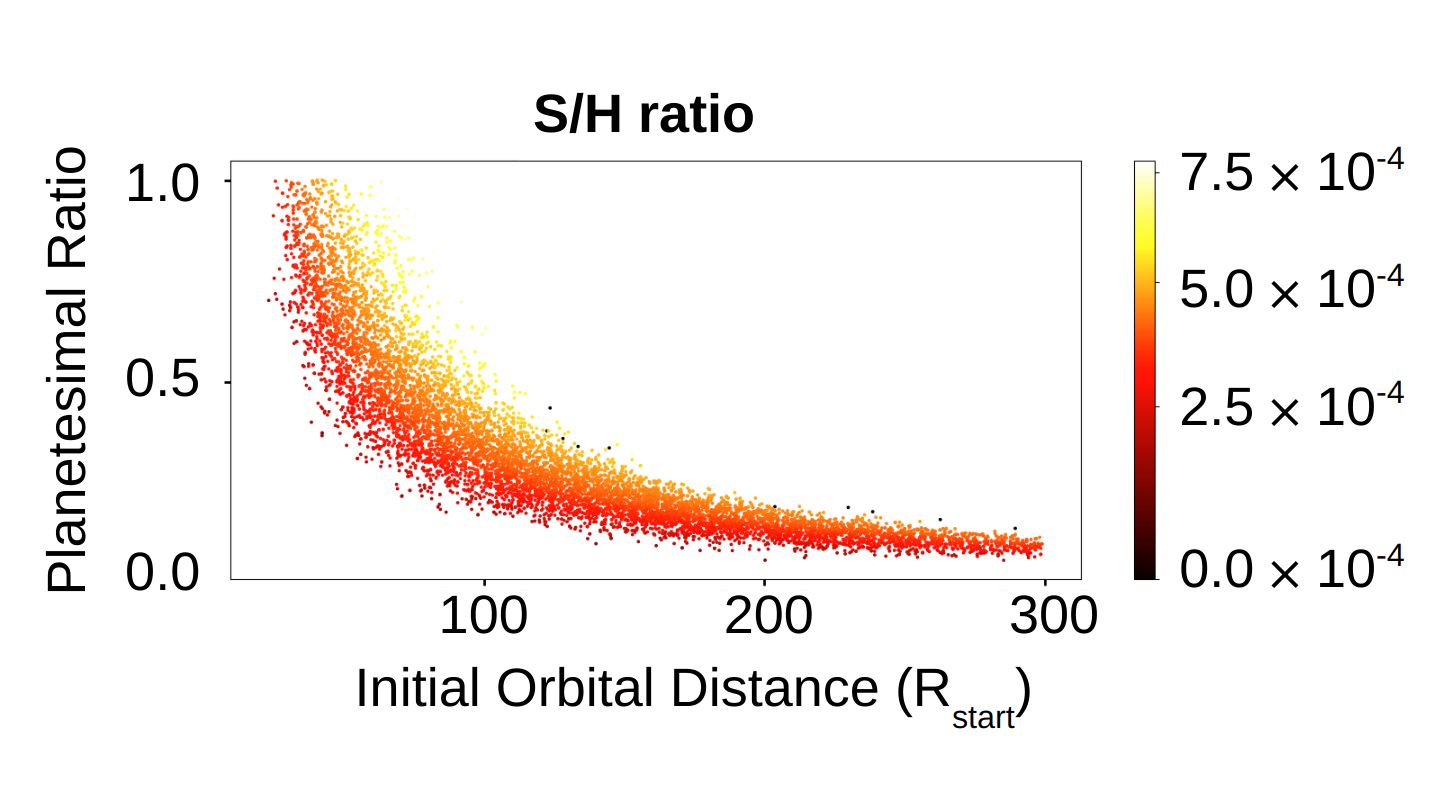}}}
     
     \caption{Correlation between the retrieved \Rstart ~and \fplanet ~ with the models' S/O and S/H ratios on the left and the right side, respectively.}
     \label{Fig: SH-SO}
\end{figure*}

In Fig.~\ref{Fig: SH-SO}, we show models from the derived posterior distribution color coded with the computed S/O ratio (left plot) and S/H ratio (right plot). We show these as a function of the planetesimal ratio and the initial orbital distance.
The correlation between these parameters suggests that by constraining the sulfur abundance we may be able to break the degeneracy between the migration distance and the planetesimal fraction.

In order to distinguish between different formation models, we compared the models with respect to how well they match the observed SO$_2$ feature by JWST. We chose several carefully selected models with different S/O ratios and S/H ratios to compare what ratio of sulfur and oxygen would best represent the observed SO$_2$ feature. 

We picked seven models from the posterior distribution to perform the photochemical calculations on. These models all have similar likelihoods. They are chosen from three groups. Two models were chosen with the goal of keeping the migration distance smaller than 100 AU and a planetesimal ratio above 0.4. The second group is chosen from models whose migration distance is between 100 to 200 AU, with a planetesimal ratio between 0.2 and 0.4. The third group is chosen from models whose migration distance is larger than 200 AU, and whose planetesimal ratio is less than 0.2.

\begin{figure}
    \centering
    \includegraphics[height=4.5cm]{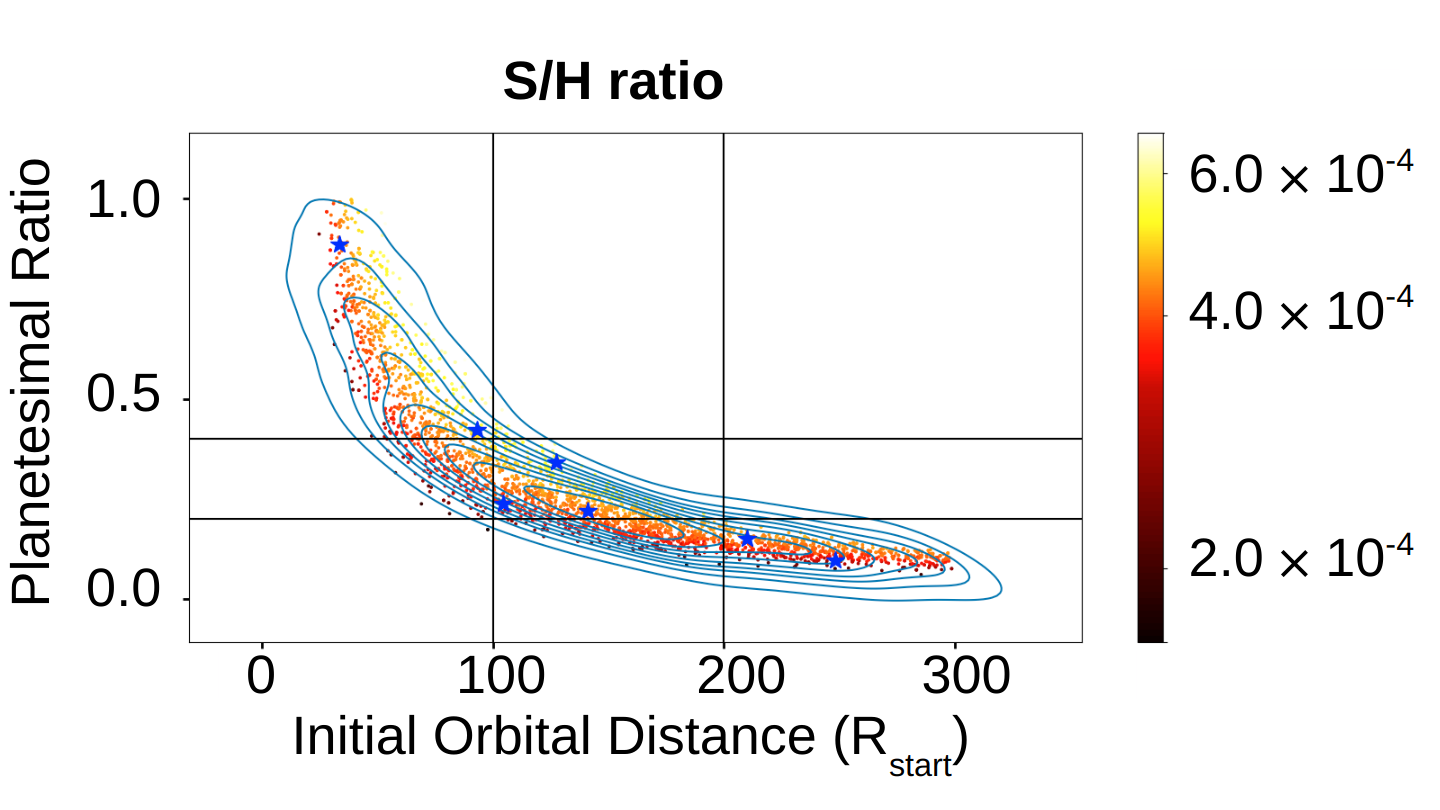}
    \caption{Location of the seven models chosen for this study (blue stars) on the \Rstart ~and \fplanet ~distribution. The vertical lines indicate migration start points of 100 AU and 200 AU. The horizontal lines indicate the arbitrary planetesimal ratios of 0.2 and 0.4.}
    \label{P3_fig: 7model}
\end{figure}

The atmospheric parameters of these models are within 1 sigma of the best-fit model. These models were chosen in a way that their cloud formation and temperature structure parameters are close to one another. The formation parameters of these planets together with their S/H and S/O are given in Table \ref{tab:7model}. Figure \ref{P3_fig: 7model} shows the planetesimal fraction and their initial orbital distance from where their Type II migration initiates.

\begin{figure}[!htbp]
    \centering
    \includegraphics[height=4.5cm]{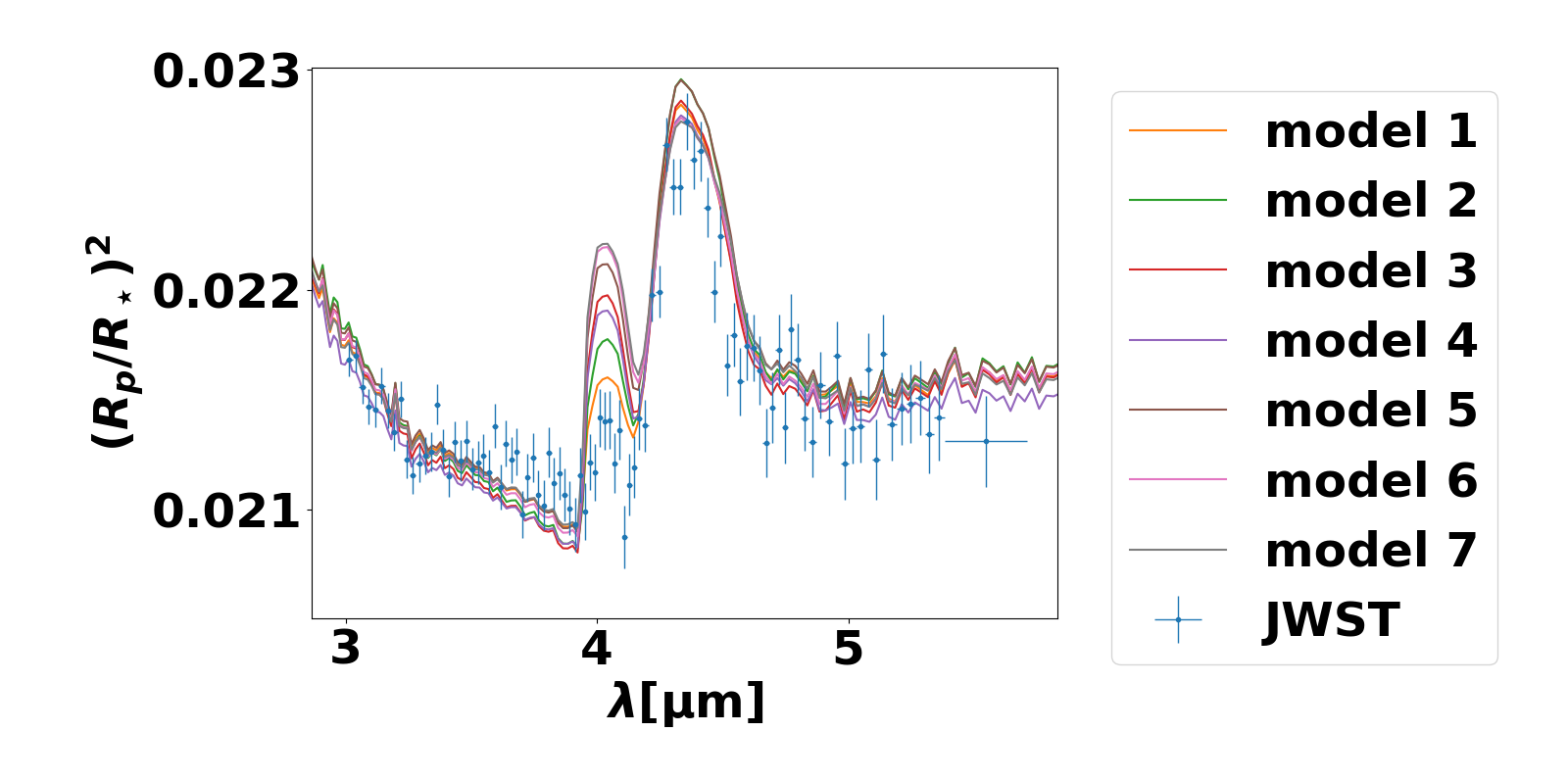}
    \caption{Comparison of the strength of the SO$_2$ and CO$_2$ features for the seven models from Fig. \ref{P3_fig: 7model}.}
    \label{fig: spectra}
\end{figure}

Figure \ref{fig: spectra} shows the modeled spectra for these seven models including the effects of photochemistry at a wavelength of 3-6 microns to compare the modeled spectra with the observation of JWST.

To understand how well these model spectra fit the observation, we fit two Gaussians to the CO$_2$ feature and the SO$_2$ feature to the modeled spectra and the observed data. Table \ref{tab:7model} shows the formation parameters of the chosen models, their solid mass contents, and their atmospheric S/H and S/O ratios for the seven models we are considering. The last column shows the ratio between the CO$_2$ feature at $4.5$ micron and SO$_2$ feature at $4.0$ micron. The first row in this table shows this ratio for the JWST observation. Looking at Fig. \ref{fig: spectra} and comparing the S/H and S/O ratios that are provided for each model in Table \ref{tab:7model}, it can be noticed that decreasing the S/H ratio the SO$_2$ feature is modeled better. On the other hand, the S/O ratio does not play a significant role in the size of the feature. By comparing the ratio between the CO$_2$ and the SO$_2$ features in the modeled spectra with that of the JWST observation, it is evident that the model with an S/H ratio of $2.27 \times 10^{-4}$ has the closest ratio to the ratio of SO$_2$ and CO$_2$ features observed by JWST. This ratio is 18.7 times larger than the solar value ($1.21\times10^{-5}$), based on the reported value from \citet{Asplund_2009}. This value corresponds to a metallicity of 1.27 in the log scale which is on the lower end of the metallicity derived from the integrated retrieval. The amount of solid material that is accreted in this model is around 14 earth masses.
\begin{table*}[]
    \centering
    \caption{Information on the seven models used in this study and the JWST observation.}
    \begin{tabular}{|c|c|c|c|c|c|c|c|c|}
    \hline
         model &\Rstart[AU]& \fplanet& \fdust&\fcarbon&Solid mass $[M_{Earth}]$&S/H&S/O&ratio of the spectral feature  \\
    \hline
    JWST& & & & & & & &0.287\\
    \hline
    1&248.28& 0.37& 0.10& 0.94&14& $2.27\times10^{-4}$&$2.72\times10^{-2}$&0.367\\
 
    2&104.43& 0.10& 0.24& 0.50& 12&$2.65\times10^{-4}$& $3.37\times10^{-2}$&0.454\\

    3&33.47&  0.85&  0.88&  0.35& 13&$3.32\times10^{-4}$& $4.10\times10^{-2}$&0.581\\
    4&210.03& 0.43& 0.15& 0.93&18&$3.47\times10^{-4}$& $2.74\times10^{-2}$&0.559\\
 
    5&140.89& 0.49& 0.22&0.28&15&$3.52\times10^{-4}$& $3.45\times10^{-2}$&0.605\\
    6&93.21& 0.72 & 0.42 & 0.28 &17&$4.40\times10^{-4}$& $4.07\times10^{-2}$&0.705\\
    7&127.38 & 0.60 & 0.34 & 0.39 & 21&$5.09\times10^{-4}$& $3.41\times10^{-2}$&0.713\\
    \hline
    \end{tabular}
    \label{tab:7model}
\end{table*}

From the above, we can conclude that the S/H ratio in the atmosphere of WASP-39b is below $2.27 \times 10^{-4}$. To find the formation and composition of planets that match the JWST observation and have S/H ratios lower than $2.27 \times 10^{-4}$, we performed an integrated retrieval and constrained the sulfur abundance using a strict prior on the S/H ratio. The prior used for the three retrievals in this study and their Bayesian evidence is reported in Table \ref{tab:my_label}. In Fig.~\ref{P3_fig: comp}, we compare the ratio of the atomic abundance of these models (blue) to the distribution that is derived for WASP-39b without such a constraint (black). This figure shows that modeled planets that have S/H lower than $2.27 \times 10^{-4}$, tend to have lower metallicities compared to those that do not have this constraint. The constraint on the S/H does not impact the models C/O, Si/O, or N/O. The models with lower metallicities are therefore better matching the SO$_2$ feature.

\begin{table}[]
    \centering
    \caption{Bayesian evidence of the three retrievals in this work.}
    \begin{tabular}{|c|c|c|}
        \hline
         Model&Prior on S/H &Bayesian evidence  \\
         \hline
         ARCiS free retrieval& no prior&1166.7\\
         ARCiS+SimAb& no prior &1169.8\\
         ARCiS+SimAb & $<2.27 \times 10^{-4}$&1168.2\\
         \hline
    \end{tabular}
    \label{tab:my_label}
\end{table}

 Figure \ref{P3_fig: form} shows the retrieved formation parameters with low S/H in blue. By comparing this distribution to the formation parameter distribution that is obtained for WASP-39b with no restriction on its sulfur abundance it is evident that planets that accrete lower mass in planetesimals generally have lower sulfur abundance. Such planets are formed by migrating from beyond the CO ice line, while accreting lower amounts of planetesimals. The distribution of \Rstart for the retrievals with a strict S/H prior is shifted towards further out in the disk compared to the retrievals without a sulfur prior. 
 
 Even though this figure shows a higher median for the \fcarbon for planets with a low sulfur abundance, the two distributions are similar enough to prevent any specific conclusions from being drawn.

\subsection{Additional spectral features}
Including photo-chemistry in the modeling of the atmosphere of WASP-39b allows us to match the SO$_2$ feature at around 4 microns in the observation of JWST. \citet{Tsai_2022} have discussed the different paths to forming SO$_2$. These authors showed that in a hydrogen-dominated atmosphere, assuming chemical equilibrium and pressures lower than 0.1 mbar, H$_2$S transforms to HS and then sulfur. In the presence of OH, sulfur can form SO and then SO$_2$. The OH used up in these reactions is formed from the photo-dissociation of H$_2$O. 

\begin{figure}[!tbp]
    \centering
    \includegraphics[height=4.5cm]{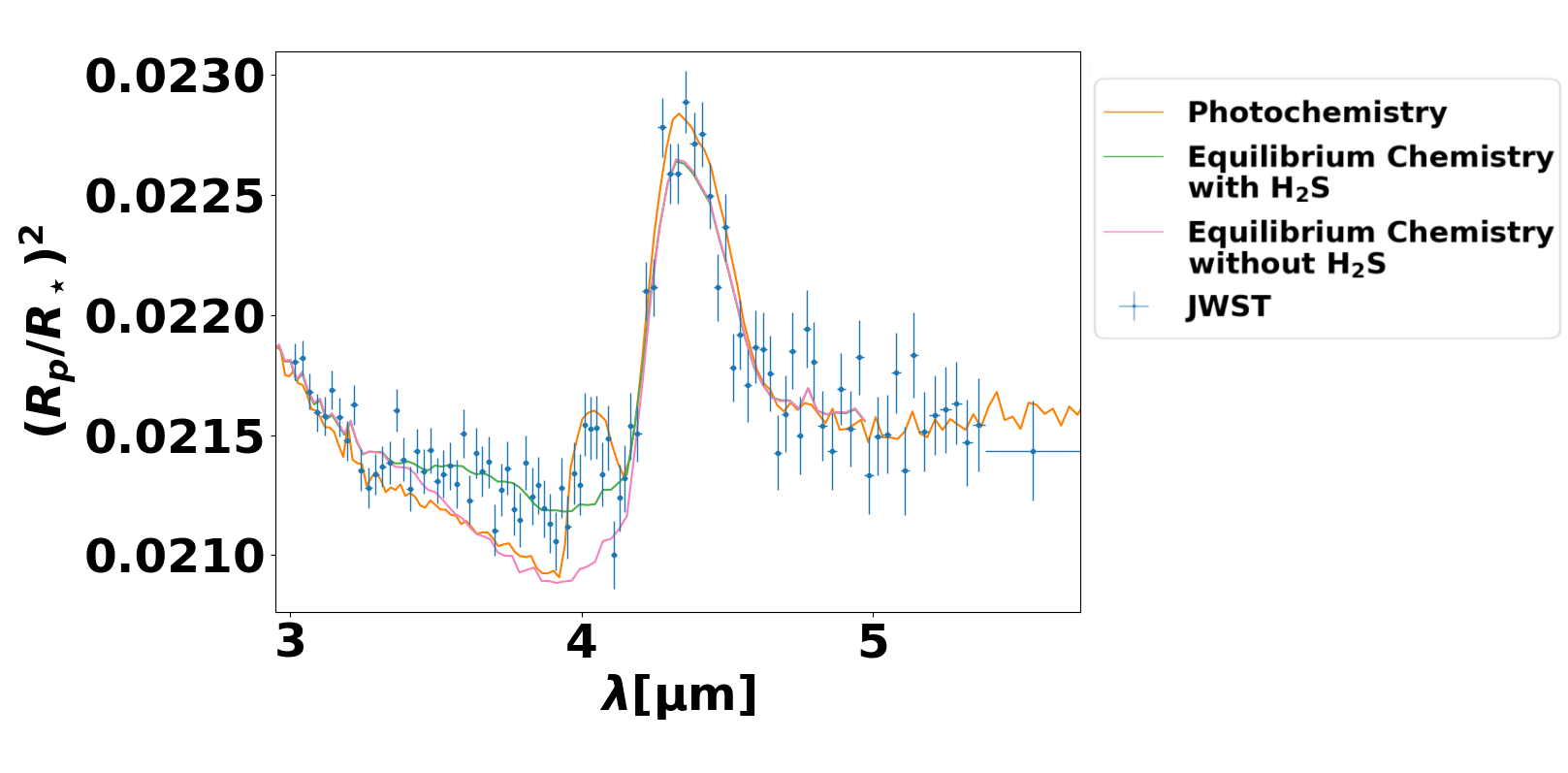}
    \caption{Comparison between the fitted spectra for three different scenarios. Pink shows the modeled spectra for an atmosphere governed by equilibrium chemistry with no H$_2$S. The green shows an atmosphere governed by equilibrium chemistry with H$_2$S. Orange shows the spectra for an atmosphere impacted by photochemistry.}
    \label{fig: H2S-SO2}
\end{figure}

 When assuming equilibrium chemistry without photochemistry, we find that H$_2$S can fit the spectral feature at around 3.7 micron. This is in agreement with the results reported in \citet{Constantinou_2023} and \citet{niraula_2023}. While other studies were unable to confirm the impact of H2S on the spectra \citep{carone_2023,Rustamkulov_2023,Alderson_2023,Feinstein_2023}. Figure \ref{fig: H2S-SO2} displays the modeled spectra for model 1, which offers the most consistency with the SO$_2$ feature (Section
\ref{sec1: S-abun}). This figure shows the modeled spectra for three scenarios, including photo-chemistry through VULCAN, equilibrium chemistry while including H$_2$S in the atmosphere, and equilibrium chemistry without including H$_2$S in the atmosphere.

This figure suggests that it is possible to fit the observation at 3.7 micron by including H$_2$S when using equilibrium chemistry. Photo-chemistry destroys H$_2$S to the extent that it is no longer visible in the spectra of the planet. The observation indicates that H$_2$S is still observable hinting that photo-chemistry does not destroy all the H$_2$S in the atmosphere. Confirmation of the H$_2$S and SO$_2$ features may provide us with information on how efficient photo-chemistry is in the atmosphere of WASP-39b. 

Since our current model does not allow for both molecules to be present in the atmosphere, we can interpret our derived sulfur abundance as a lower limit.

\section{Discussion}
\label{sec: discuss}

Below we discuss the implications and limitations of the study presented.

\subsection{Formation of WASP-39b}

Studies such as \citet{Mordasini_2016} have shown that the main source of solid accretion for planets with super-solar metallicities are planetesimals. This is in agreement with the planet formation formalism we use in this paper. The amount of planetesimals accreted onto the planet depends on the migration distance and the planetesimal ratio. The planetesimal ratio parameterizes the planet's efficiency in accreting planetesimals as well as the availability of planetesimals in the disk. This explains the strong degeneracy between the planetesimal fraction and the initial orbital distance from where the planet initiates its migration.
Within the framework of SimAb, WASP-39b should have initiated its Type II migration from beyond the CO$_2$ ice line. This is a direct constraint from the high metallicity and the solar C/O ratio.

The retrieval method we used to study the formation parameters of WASP-39b is not able to constrain the dust grain fraction. This can be understood in the context of paper I{,} where we have shown that the accretion of dust grains does not influence the metallicity of the planets with super-solar metallicities. The results of this study suggest that the metallicity of WASP-39b is more than ten times solar, which is in agreement with previous studies such as \citet{Rustamkulov_2023,carone_2023, Min_2020}. This means that it is not possible to retrieve the dust grain fraction given the model we have used in this work.

In paper I, we predict that the C/O ratio would help disentangle the model degeneracy between the planetesimal fraction and the initial orbital distance. However, Fig. \ref{P3_fig: comp} shows that there are not much constraints on the C/O ratio obtained from the free retrieval. This means that the constraint that is seen for the atmospheric C/O ratio when using the integrated ARCiS-SimAb models is most likely due to the prior that is enforced by the model. With respect to paper I, we added a fourth parameter, the carbon fraction, which is an additional complexity that makes this degeneracy harder to resolve.

The carbon fraction at the soot line mainly impacts the carbon abundance in the atmosphere of planets with respect to other elements such as oxygen. Increasing the carbon fraction at the soot line allows for higher C/O ratios in the atmosphere of planets with super-solar metallicities. Hence, it is possible for planets to acquire C/O ratios closer to solar when they are formed within the CO ice line in a disk with a larger carbon fraction at the soot line. We find that with the current observations, it is not possible to put a strong constraint on the carbon fraction at the soot line.

\subsection{Atmospheric composition}

When retrieving the atmospheric composition, ARCiS allows for a wide range of values for the metallicity, N/O ratio, Si/O ratio, and C/O ratio. Additionally, these parameters are independent of one another. Therefore, the ratios that are retrieved for the atomic abundance of the planet's atmosphere using ARCiS denote a general atmospheric composition, uninformed by planet formation constraints. On the other hand, by applying SimAb{,} we derive the atmospheric abundances based on the formation parameters. This means that the atmospheric abundances are dependent on one another, while the formation parameters are independent of one another. Consistency between the ratios of the atomic abundances of the atmosphere that is retrieved using ARCiS free retrieval and those derived through ARCiS-SimAb integrated method shows that SimAb predictions are compatible with the retrieved atmospheric composition of WASP-39b, governed by its atmospheric observations.

\subsection{Constraining formation by using sulfur abundance}

In this study, we used the SO$_2$ feature at 4 microns to constrain the S/H abundance. We show that the sulfur abundance puts further constrain on the metallicity of the atmosphere of the planet. This provides us with an estimate of the accreted mass from the solid phase, including planetesimals and dust grains, to be $11.73^{+1.64}_{-1.60}$ earth masses. It is important to mention that this mass is less than the total mass in heavy elements accreted on the planet. The gas composition in the disk includes heavy elements ,contributing to the metallicity of the planet, such as oxygen and carbon in the form of volatiles, which are not included in the reported mass of the accreted solid phase. The mass of the total heavy element can easily be calculated from the metallicity of the planet. Even though the retrieved planetesimal ratio and initial orbital distance are model dependent, the amount of heavy element mass in the atmosphere is directly related to the observation of the atmosphere and its metallicity. By considering the amount of accreted volatiles, the solid accretion during formation should be consistent between models.

Our results show that by putting a stricter prior on the sulfur abundance, the retrieval results in a formation path that initiates its migration further out by accreting fewer planetesimals compared to when there is no prior on the sulfur abundance. Furthermore, such models result in lower metallicities. By lowering the planetesimal ratio, the lower metallicity is easier achieved. Increasing the initial orbital distance from where the planet migrates serves two purposes. First, by accreting material from beyond the CO ice line, the forming planet gains access to a gas phase primarily formed of hydrogen and helium. Second, the majority of the solid mass that is accreted from beyond the CO ice line mainly consists of volatile oxygen and carbon. Hence, solids in this region have lower sulfur abundance per unit of mass. The slight increase in carbon fraction at the soot line may play a similar role, by increasing the carbon abundance per unit of mass. In the future, retrievals using photochemical models might provide a more consistent coupling between the sulfur abundance and observed features from sulfur-bearing species, tightening the constraint on planet formation.

\subsection{Post formation atmospheric evolution}

Over time, the atmospheres of close-in planets lose part of their hydrogen abundance and become more metallic \citep[e.g., through photoevaporation][]{Locci_2019}. This causes the atmospheric metallicity derived through atmospheric observation to represent an upper limit for the planet's metallicity immediately after formation. This is true for the ratio of heavy elements compared to hydrogen, such as the S/H ratio which is of interest in this study. However, the ratio of heavy elements compared to one another, namely the C/O ratio, is expected to stay intact during the life span of a planet. Therefore, the ratios of elements heavier than H and He provide more robust tracers of the formation history.

\citet{Louca_2023} have shown that the metallicity of WASP-39b, considering that atmospheric loss is at most enhanced by a factor of 1.24 $\times$ the initial metallicity. This indicates that the high metallicity of the planet is mainly associated with its formation and the impact of the metallicity enhancement due to the loss of atmosphere is negligible.

\subsection{The impact of cloud formation on the sulfur abundance}
{In our retrieval framework, we include cloud formation self-consistently. However, we did not couple cloud formation to the photochemical computations, and the presented spectra from the photochemical computations are cloud-free.} We expect a marginal impact on the SO$_2$ feature caused by cloud formation. The atmospheric temperature of WASP-39b is above 700K which is higher than the condensation temperature of the main solid sulfur reservoir FeS. As discussed in \citet{Tsai_2022}, other sulfur-bearing species that may condense at such temperatures would not impact the abundance of SO$_2$ by much. Furthermore, \citet{Polman_2022} also have found that clouds above pressures of 10$^{-3}$ bar should not affect the SO$_2$ feature, due to SO$_2$ only being abundant at low pressures.

The abundance of CO$_2$, on the other hand, can be affected by the formation of clouds. Oxygen is one of the main elements in clouds{;} therefore, cloud formation can impact the C/O ratio and the abundance of CO$_2$. Hence, by ignoring cloud formation, we may overestimate the CO$_2$ feature in the spectra for the models we chose. However, Fig. \ref{fig: H2S-SO2} shows negligible differences between the CO$_2$ feature modeled using photochemistry and equilibrium chemistry. As the models using equilibrium chemistry also include the effects of cloud formation, this figure suggests that the cloud formation effect on the CO$_2$ feature is negligible for the results of this study.

\subsection{Other limitations of this work}
\label{sec1: lim-work}

Within VULCAN, we used a photospheric fit of HD 209458 combined with the remaining part of the spectrum of HD 189733. The high-energy spectral energy distribution of WASP-39 has not been measured, requiring us to use a combination of different spectra. HD 209458 is a G-type star, just as WASP-39, although HD 209458 is slightly hotter. The spectrum of HD 189733 is less good of a fit with WASP-39, due to HD 189733 being significantly younger, thus having more chromospheric activity. This method severely underestimates the stellar flux in the optical range. However, this is not relevant for our goals as this wavelength range does not affect photochemistry. Both using the photospheric fit of a hotter planet and using a star with more chromospheric activity for the lower wavelengths contribute to an overestimation of the radiation incident on WASP-39b. While this is not ideal, \citet{Polman_2022} have suggested that the impact on the SO$_2$ feature is small. They reported a change in feature size of $\sim$10\% between stellar flux spectra of HD 189733 and HD 209458 for strong features. The currently used spectrum and the actual spectrum of WASP-39b show less of a variation, suggesting that differences in the feature size of SO$_2$ should be less than 10\%. 

The photochemistry model that we used predicts that in the atmosphere of WASP-39, most of the sulfur forms SO$_2$, whereas using equilibrium chemistry predicts the formation of H$_2$S. We show that part of the sulfur possibly stays in the form of H$_2$S which causes a spectral feature at 3.7 micron. If this H$_2$S feature is confirmed, this might indicate that photochemical conversion into SO$_2$ is less efficient than currently predicted.

One of the limiting factors that can impact the results is that SimAb assumes the chemistry of the disk does not evolve as the planet is forming. However \citet{Notsu_2020} have shown that by assuming a chemically evolving disk, the C/O ratio in the disk varies compared to what is simulated in this study. Furthermore, \citet{Eistrup_2018} have shown that chemical evolution in the disk is more drastic for older disks, specifically for planets that pass ice lines during their formations, thus making them relevant to this study. However, a chemically evolving disk would not impact the amount of solid phase accretion of the planet, based on the other assumptions in this model and, consequently, its sulfur content. Nonetheless, it would impact the overall metallicity of the planet. Therefore, it is complicated to predict the impact and the significance of such an assumption. Constructing a more realistic chemical composition for the disk while the planet is forming is an important future step that is beyond the scope of this work.

In this study we used an equilibrium chemistry to retrieve the atmosphere of WASP-39b. However, \citet{Kawashima_2021} have shown that disequilibrium chemistry is the more likely scenario to explain the atmosphere of WASP-39b. These authors showed that disequilibrium chemistry results in a lower metallicity for the atmosphere of the planet compared to when assuming equilibrium chemistry. This suggests that using disequilibrium chemistry could result in a different formation history for WASP-39b. However, as is discussed in \citet{Tsai_2021}, the temperature-pressure profile of the planet does not vary enough in these cases to affect the photochemistry results. Therefore, we expect a similar constraint on the atmospheric S/H ratio of the planet and, hence, on the formation parameters derived from this value.

\section{Conclusion}
\label{sec: conclusion}
In this study we look at the formation history of WASP-39b, using the transit observations done by JWST, HST, and Spitzer. We use a novel approach to study the formation of this planet by integrating the formation model SimAb in the atmospheric retrieval code ARCiS. Then, SimAb replaces the parameters that predict the atmospheric atomic composition of the planet. This approach allows us to study the formation of the planet directly from the observations. We show that this approach does not result in a significantly different temperature and pressure profile as well as cloud formation parameters for the planet compared to ARCiS free retrieval. 

Our formation study, when assuming equilibrium chemistry in the atmosphere of WASP-39b, suggests that this planet initiated its Type II migration from beyond the CO$_2$ ice line with a higher possibility for formation within the CO ice line. Although there is no constraint on the amount of dust grain accretion during the formation of the planet, a lower limit of $0.14$ is obtained for the planetesimal ratio.

By including photochemistry to obtain the sulfur abundance using the SO$_2$ feature at around 4 micron, we show that the S/H ratio should be less than $2.27 \times 10^{-4}$. Planets that form at farther distances, beyond the CO$_2$ ice line and CO ice line, are more likely to have a sulfur abundance less than this value. Including this prior results in a planetesimal ratio of $0.13^{+0.12}_{-0.04}$. This value is more constrained compared to when there is no prior on the sulfur abundance. Furthermore, the metallicity that is obtained when including a restricted prior on the sulfur abundance, is lower than what is predicted from the ARCiS free retrieval. These planets accrete fewer planetesimals, hence the lower retrieved planetesimal ratio. These results demonstrate the power of sulfur to provide additional constraints on planet formation scenarios and stress the importance of including additional abundance constraints beyond carbon and oxygen.

When using ARCiS forward modeling in three scenarios, equilibrium chemistry with no H$_2$S, equilibrium chemistry with H$_2$S, and photochemistry, we see that these spectra are different. The model where we assume equilibrium chemistry and include H$_2$S suggests that a H$_2$S feature is present at around 3.7 microns. This feature is not visible in the other two models, while the transiting spectra are not well matched with the modeled spectra. On the other hand, including photochemistry in the forward model suggests that SO$_2$ feature at around 4 microns is present which is not present in the other two models. Confirmation of both these features implies that sulfur is not solely in SO$_2$; it is possible that part of the sulfur abundance has remained in the H$_2$S. Further work is required to include photochemistry in the retrieval models and allows for balancing the presence of H$_2$S and SO$_2$ in the atmosphere modelings. Such a model allows us to study the importance of photochemistry in the atmosphere of planets as well providing a more accurate estimate of sulfur abundance which is then used to understand the formation of planets.

\section*{Acknowledgments}
The contribution of J.P. has been carried out within the framework of the NCCR PlanetS supported by the Swiss National Science Foundation under grants 51NF40\_182901 and 51NF40\_205606.

        \newpage

        \bibliographystyle{aa}
        \bibliography{reference} 

\end{document}